# The first principles calculations of the atomic and electronic structure of cubic and orthorhombic LaMnO$_3$ surfaces

Yu. A. Mastrikov[1,2], E. A. Kotomin[1,2], E. Heifets[1], and J. Maier[1]

[1]Max Planck Institute for Solid State Research, Heisenbergstr. 1, D-70569, Stuttgart, Germany
[2]Institute for Solid State Physics, University of Latvia, Kengaraga str. 8, Riga LV-1063, Latvia

**Abstract.** Combining GGA-plane wave approach as implemented into the VASP - 4.6.19 computer code with a slab model, we studied in detail the atomic and electronic structure of the LaMnO$_3$ surfaces, in both cubic and orthorhombic phases. The results obtained are compared with similar studies for other ABO$_3$-perovskites.

## 1. Introduction

The ABO$_3$-type perovskite manganites (B=Mn) are important functional materials with numerous high-tech applications [1]. Some of these applications, e.g. cathodes of solid oxide fuel cells (SOFC) [2-4], need understanding and control of LaMnO$_3$ (LMO) *surface properties*. LMO has a cubic structure above 750 K, below this temperature the structure is orthorhombic, with four formula units in a primitive cell. Due to the high spin state of Mn$^{3+}$ ion, the LMO reveals several magnetic configurations. Below $T_N$=140 K the A-type antiferromagnetic (AF) phase is the ground state. This corresponds to the ferromagnetic coupling in the basal *ab* (*xy*) plane combined with an antiferromagnetic coupling in the *c* (*z*) direction (in the Pbnm setting). There exist also other magnetic states: the FM, GAF and CAF: In particular, the FM corresponds to a fully ferromagnetic material, in GAF all the spins are antiferromagnetically coupled to their nearest neighbors, and in CAF cell the spins are antiferromagnetically coupled in the basal plane and ferromagnetically between the planes (along the *c* axis).

Several first-principle studies of the LMO surfaces were undertaken recently ([5] and the references therein). However, these studies were limited to the cubic structure, often for unrelaxed surfaces. In this paper, we compare properties of three cubic and orthorhombic LMO surfaces. Section II deals with computational details, Section III describes results on the atomic and electronic structure, whereas the. Conclusions are presented in Section IV.

## 2. Computational Details

In this study, we employed the VASP 4.6.19 computer code [6] with Perdew-Wang 91 GGA nonlocal exchange-correlation functional and the basis set of plane waves (PW). More details are given in Ref.[5]. We demonstrated therein that the VASP code reproduces well the LMO lattice constants (in both phases with accuracy < 1%) as well as the cohesive energy and magnetic coupling constants. The effective atomic charges are calculated using the Bader (topological) analysis [7].

In a surface modelling, we used a slab model infinite in the two ($x$, $y$) dimensions and containing a finite number of planes along the $z$ axis normal to the surface. The orthorhombic surface unit cell has area twice larger as that for a cubic unit cell.

We started with the (001) surface, which has $MnO_2$/LaO/$MnO_2$/… sequence of the oppositely charged planes (±1 $e$, assuming formal ionic charges: $Mn^{3+}$, $La^{3+}$, $O^{2-}$). Table 1 schematically shows two types of slabs used. The *stoichiometric* cubic 8-plane (with an even number of planes, in general) slab (a) consists of a four LMO formula units (each pair of neighbouring planes gives a formula unit per unit area of surface). Its potential shortcoming is that this slab has a dipole moment normal to the surface, due to summation of the electrostatic fields of all planes [8, 9]. Such a slab could be unstable with respect to reconstruction or defect formation. Alternatively, the dipole moment can be cancelled by the charge redistribution near the surface (to be discussed below). We need also a study of such slabs for further thermodynamical analysis of surface stability.

To get rid of the dipole moment, *symmetrical* slabs (Table 1 b, c) are often used. Such slabs have an odd number of planes (typically 7 in our calculations) and the *same* terminations on both sides, e.g. $MnO_2$ or LaO. The later allows to decouple the surfaces with different terminations and to study these individually. Such slabs are not stoichiometric, but two symmetrical slabs with different terminations as put together contain an integer number of bulk unit cells. In particular, for the slab in Table 1a $n = 4$.

Based on these considerations, two relations are used for the *cleavage energy*. For stoichiometric slabs the average energy over two terminations

$$E_{cl} = \frac{1}{2S} \cdot (E_{slab} - n \cdot E_{bulk}) , \qquad (1)$$

where $S$ is an unit area of surface, $E_{slab}$ the total energy of the slab, $n$ corresponding number of the bulk units, $E_{bulk}$ total energy of the bulk unit. As a reference, the energies of the bulk unit cell of the corresponding structure and magnetic ordering were taken.
For non-stoichiometric slabs one gets, respectively,

$$E_{cl} = \frac{1}{4S} \cdot (E_{t1} + E_{t2} - n \cdot E_{bulk}) , \qquad (2)$$

where $E_{t1}$ and $E_{t2}$ denote the total energies of the slabs with odd number of planes with complementary (LaO- and $MnO_2$-) terminations; $n = 7$ for slabs (b, c) in Table 1.
When ions are allowed to relax to the minimum of the total energy, we call this *relaxed* cleavage energy. A large vacuum gap of 15.8 Å between slabs periodically repeated along the $z$ axis was used here, in order to prevent an interaction between the two surfaces through the vacuum region.

## 3. Main Results

### 3.1. The (001) Surface

The calculated (001) cleavage energy (Table 2) is quite low and close to that for $SrTiO_3$ (001), viz. 1.2 $e$V [10]. The results saturate at slab thickness of 7-8 planes. The two different ways of cleavage energy calculations, using slabs with odd and even number of planes, Eq. (1) and (2), give similar results.

Thus, the 7- and 8- plane slabs were chosen for further calculations of the cleavage energy for slabs with different magnetic ordering (Table 3). The calculations on 7-plane slabs were performed with both fixed and optimized total magnetic moments of a slab. That is, we studied how the magnetic ordering of a slab affects the results. In some slab calculations the total magnetic moment was fixed (0 for all AF states and $4\mu_B$ per each $Mn^{3+}$ ion in the FM state), whereas in other calculations, the magnetic moment was self-consistently optimized.

As follows from Table 3, the considerable effect of magnetic moment optimization on the cleavage energy is seen only in the FM state with a relaxed geometry for both cubic and orthorhombic slabs (changed by 0.28 and 0.21 $eV/a_0^2$), whereas for AFM slabs this effect is negligible ($< 0.06$ $eV/a_0^2$). The lowest cleavage energy of 0.61 $e$V is found for the G-AF magnetic configuration. For a comparison with a cubic structure, we calculated also the cleavage energy for the orthorhombic slab in the FM state. This energy exceeds by 0.14-0.21 $e$V that for a cubic slab in the same magnetic state. Note also that the FM slabs, both cubic and orthorhombic, show a cleavage energy slightly larger than that for the AFM slabs.

A full slab *geometry optimization* was performed in the present calculations. For most slabs in a cubic phase, the $z$ coordinate had to be varied only, due to high slab symmetry. In the orthorhombic FM slab only some minor atomic displacements in the planes parallel to the surface took place. The largest atomic displacements were observed along the $z$ axis, similarly to those in the cubic slabs.

Changes in the slab geometry caused by its relaxation could be expressed in terms of the interplanar and Mn-O relative distances in the $MnO_6$ octahedra along the [001] direction perpendicular to the surface. In the cubic phase of LMO bulk, the optimized distance between the nearest $MnO_2$ and LaO planes is 1.95 Å (left side of Fig. 1, a,b). However, near the two surface terminations these planes are splitted-off (right side of Fig.1.a,b). The splitting of the top plane is called *rumpling*. As a result, we obtained four most important different (interplanar) distances in the relaxed cubic slabs: the top Mn-$O_2$ plane rumpling, $O_2$-La distance, La-O plane splitting, and O-Mn distance between planes II and III (Figure 1a, b). The rumpling of the $MnO_2$ and LaO surface planes in the cubic slabs is about 0.19 Å and 0.44 Å, respectively (Table 4), which could be checked by means of LEED experiments. These distances vary, depending on the proximity to the surface or particular termination, what can be easily identified in the 8-plane slabs (Table 4). The Mn-$O_2$ plane splitting and Mn-O interplanar separation distance monotonically decreases and increases, respectively, when going along the [001] direction from the $MnO_2$- towards the LaO-termination (planes I to VIII in Table 4). For both terminations the $O_2$-La distance and La-O plane splitting increases and decreases from the surface towards the slab centre, respectively. Notice that Mn ions move always above O ions from the same plane whereas La ions move below O in LaO- , but above O ions in the $MnO_2$- termination.

In the orthorhombic bulk the $MnO_2$ plane is split-off into three subplanes as shown in Fig. 1. c,d (left side) whereas near the surface LaO plane is additionally split-off (Fig. 1.c,d, right side). The O-Mn distance along the [001] direction in the orthorhombic cell is 1.97 Å. The interplanar distances calculated for the orthorhombic bulk are (Fig. 1.c,d):

LaO (I) –O(II) 1.66 Å, whereas the splitting of plane II: O-Mn 0.31 Å. All calculated slabs with different magnetic orderings for the corresponding terminations demonstrate similar atomic displacements. There is a small difference between the relaxation in the odd-plane slabs near the mirror plane and that in the central planes of the even-layer slabs, caused by symmetry of the slabs. The $MnO_2$ plane splitting into three subplanes mentioned above arises due to the tilting of the $MnO_6$ octahedra. The sequence of the subplanes here is the same as in the bulk (O-Mn-O), except for the terminating $MnO_2$ plane. The Mn-subplane moves threin outwards the slab centre, leaving both top O subplanes below (Figure 1c). The outward move of Mn ions from the $MnO_2$ plane on the $MnO_2$-terminated surfaces was also observed for the cubic slabs. Instead, for the LaO-termination, La atoms are covered by O subplane. This is true for both asymmetrical and symmetrical (001) slabs.

Lastly, Table 5 summarizes the Mn-O *distance variation* in the $MnO_6$ octahedra along the *z* axis. As one can see, the upper side of the $MnO_6$ octahedra in stoichiometric slab is regularly expanded in the direction of the LaO termination, but the lower part compressed, as compared to a perfect bulk octahedron. This effect depends again very weakly on the magnetic state and slab stoichiometry.

In order to illustrate the *charge density redistribution* in the stoichiometric 8-plane slab, the total and difference density map (with respect to the bulk) were plotted in Fig.2. The total density maps clearly show the *zig-zag-type* wavy orthorhombic structure of slabs. The difference maps demonstrate, in its turn, that only near-surface metal atoms are visibly perturbed (Mn atoms considerably polarized) and that there is no interaction between two terminating surfaces of a slab. A comparison with similar calculations for the symmetric 7-plane slab shows that in spite of the different stoichiometry of the 7- and 8-plane slabs, this perturbation is very similar for the corresponding terminations of both slabs. This conclusion is important for the modeling of O an interaction with LMO surfaces (in progress).

To shed more light on the electronic density redistribution, the Bader *effective atomic charges* were calculated for 7- and 8- plane slabs in different magnetic states. First of all, the effective charges vary very slightly with the magnetic state. Second, the surface-induced perturbation is not restricted to the first plane like in semiconducting $SrTiO_3$ [11] but for both terminations the perturbation remains considerable in several planes below the surface – in line with the results for the stoichiometric slab and previous HF calculations [12, 13].

The effective atomic charges in the orthorhombic phase are similar to those in cubic slabs in the same FM state. However, these charges are much closer to those in the orthorhombic bulk than in a cubic phase.

A comparison of the two nonstoichiometric slabs (with $MnO_2$ and LaO terminations on both sides Table 6 and Table 7, respectively) shows that the deviations of the *total* slab charge with respect to the bulk (Table 6d) are equal but have opposite signs (+/- 0.66*e* for the FM state). The deviation of this charge from +/- 1*e*, which is expected for the formal ionic charges ($O^{2-}$, $La^{3+}$, $Mn^{3+}$), is due to the partial Mn-O bond covalency.

A similar analysis was performed for stoichiometric slabs. The charge of LaO surface is reduced compared to the bulk, which tends to reduce the dipole moment of the slab, according to electrostatic arguments [8]. On the other hand, the charge of the $MnO_2$ surface practically does not change; moreover, larger perturbations are observed in the deeper, third layer from the surface plane (unlike the LaO-termination). In general, metal ions (La and Mn) show larger charge changes than O atoms. The calculated charges depend weakly on the slab magnetic states.

*3.2. The (110) Surfaces*

The atomic and electronic structure of the polar (110) $LaMnO_3$ surfaces was calculated for the FM configuration and both cubic and orthorhombic phases. Similarly, both the 8-plane stoichiometric asymmetrical slabs $O_2$/LaMnO/... LaMnO and two types of 7-plane nonstoichiometric but symmetrical slabs without dipole moments ($O_2$/LaMnO...$O_2$ and LaMnO/$O_2$...LaMnO) were modeled as shown in Table 8 and Fig.3.

The (110) $O_2$-terminated 7-plane slab could be made stoichiometric by removing from both terminating surfaces half of O ions, i.e. leaving one O ion instead of the two in the surface unit cell. This procedure is widely used for stabilization of the polar (110) oxide surfaces, e.g. for $SrTiO_3$ (110) [9, 14]. In our study, we used 2x1 extended surface unit cell and thus removed half of O ions from nearest surface cells in the zig-zag way, in order to give the surface more degrees of freedom for a further relaxation [15]. In calculations of the cleavage energy, Eqs. (1) and (2) were used.

As follows from Table 9, in all three cases, the (110) cleavage energy is larger than that for the above - discussed (001) surface. The same conclusion was drawn in the HF calculations [12]. Second, the VASP-calculated cleavage energies for 7- and 8-plane slabs practically coincide. This shows that the macroscopic dipole moment of a stoichiometric slab plays no essential role in the present calculations. Very important is that the calculated cleavage energy with a half of O atoms removed is considerably lower than the defectless surface. In other words: the removal of a half of O atoms from the polar (110) surface has indeed a strong *stabilizing effect*. A similar conclusion was drawn earlier based on Shell Model and unrelaxed HF calculations [12, 15].

We have compared the atomic relaxations for 7- and 8-plane slabs in Table 10. Unlike the (001) surface, atoms in $O_2$-planes now experience (even without defects) in-plane displacements along the *y* axis. Moreover, we observe large surface La displacements inwards the slab center (6-7% of $a_0$), whereas Mn and O ions move in the opposite direction. As a result, we predict that this surface has to exhibit very large rumpling, which could be checked by means of LEED experiments. Considerable difference between atomic displacements in 7- and 8-plane slabs indicates that these slabs are still not thick enough, what is in line with our similar conclusions for the (110) $SrTiO_3$ [16]. This is supported by the fact that atomic displacements are large even in the slab centre.

Analysis of the effective charges demonstrates that the LaMnO-terminated surface is strongly negatively charged with respect to the bulk (-1.14 *e*), whereas the second plane charge is already close to the bulk. The complementary $O_2$–terminated surface is positively charged (0.79 *e*), and deeper perturbed. This effect is observed in both stoichiometric and nonstoichiometric slabs.

Lastly, atomic relaxation on the O-terminated (110) surfaces containing O vacancies in the two possible configurations (Fig.4) are shown in Table 11. The defect creation on the surface induces considerable (~10%) in - plane and inward atomic displacements.

*3.3 The (111) surface*

We calculated also the cleavage energies for the 7- and 8-plane (111) surfaces with Mn- and LaO$_3$- terminations (Table 12). The main conclusion is that this energy is much larger than that for the (001) surface. The energy estimates based on Eqs.(1) and (2) are very close. Introduction of surface defects (Mn-vacancies) stabilizes the surface but not enough to compete with the (001) surface energy.

**4. Conclusions**

1. Based on our calculations, we predict large rumpling of the LMO (001) surface. This could be checked by means of the LEED experiments.
2. The calculated Mn-O distances along the *z* axis show that the upper parts of MnO$_6$ octahedra (with respect to the LaO-termination) are regularly expanded whereas the lower parts compressed as compated to the perfect bulk MnO$_6$ octahedron.
3. There is a considerable electronic density redistribution near the surface which could affect atomic and molecular adsorption (simulations in progress). The calculated effective atomic charges weakly depend on the magnetic structure and slab stoichiometry.
4. The (110) surface is stabilized by a partial removal of oxygen ions from the surface, similarly to SrTiO$_3$ [14], revealing the smallest cleavage energy among studied. This surface demonstrates even larger rumpling, more deep perturbation with respect to the bulk and in-plane ionic displacements, unlike the (001) surface. The cleavage energies for the defectless (110) and (111) surfaces are also larger than that for the (001).
5. These results are used in the thermodynamical analysis of the LMO surface stability under different partial oxygen gas pressures and finite temperatures (in progress).

**Acknowledgements**

This study was partly supported by the Latvian National Program on Material Science.

Table 1. Plane sequence for the (001) 8-plane (a) and 7-plane $MnO_2$-(b) and LaO-(c) terminated slabs.

| a) | b) | c) |
|---|---|---|
| $MnO_2$ | $MnO_2$ | |
| LaO | LaO | LaO |
| $MnO_2$ | $MnO_2$ | $MnO_2$ |
| LaO | LaO | LaO |
| $MnO_2$ | $MnO_2$ | $MnO_2$ |
| LaO | LaO | LaO |
| $MnO_2$ | $MnO_2$ | $MnO_2$ |
| LaO | | LaO |

Table 2. Cleavage (relaxed and unrelaxed) energies ($eV/a_0^2$) for the (001) cubic slabs in the FM state. For better comparison, slabs with even and odd number of planes are grouped in two different pairs of columns. The relevant energies were calculated using Eqs. (1) and (2), respectively.

| N. of planes | unrelaxed | relaxed | unrelaxed | relaxed |
|---|---|---|---|---|
| 4 | 1.64 | 0.76 | | |
| 6 | 1.66 | 0.82 | | |
| 7 | | | 1.58 | 1.01 |
| 8 | 1.66 | 0.82 | | |
| 9 | | | 1.59 | 0.97 |
| 10 | 1.66 | 0.83 | | |
| 11 | | | 1.60 | 0.96 |
| 12 | 1.67 | 0.84 | | |
| 13 | | | 1.61 | 0.95 |

Table 3. Cleavage energies of the (001) surface ($eV/a_0^2$) for cubic slabs in different magnetic configurations.

| slab | 7-plane | | | | 8-plane | |
|---|---|---|---|---|---|---|
| total magnetic moment | fixed | | optimised | | fixed | |
| | unrelaxed | relaxed | unrelaxed | relaxed | unrelaxede | relaxed |
| FM | 1.60 | 1.25 | 1.55 | 0.97 | 1.61 | 0.87 |
| AAF | 1.53 | 0.95 | 1.53 | 0.94 | 1.51 | 0.77 |
| CAF | 1.46 | 1.01 | 1.43 | 0.95 | 1.46 | 0.78 |
| GAF | 1.34 | 0.76 | 1.34 | 0.75 | 1.43 | 0.61 |
| FM* | 1.84 | 1.39 | 1.77 | 1.18 | 1.77 | 1.13 |

* orthorhombic slab

Table 4. The interplanar distances (Å) in the relaxed 8-plane cubic slab (Fig.1).

| plane | subplane | FM | AAF | CAF | GAF | plane | subplane | FM* |
|---|---|---|---|---|---|---|---|---|
| | | | | | | I | Mn-O | 0.08 |
| I | Mn-$O_2$* | 0.19 | 0.19 | 0.16 | 0.16 | I | O-O | 0.17 |
| I-II | $O_2$-La | 1.54 | 1.55 | 1.53 | 1.54 | I-II | O-La | 1.47 |
| II | La-O* | 0.39 | 0.39 | 0.42 | 0.43 | II | La-O | 0.37 |
| II-III | O-Mn | 1.88 | 1.88 | 1.87 | 1.87 | II-III | O-O | 1.68 |
| III | Mn-$O_2$* | 0.15 | 0.15 | 0.16 | 0.16 | III | O-Mn | 0.16 |
| | | | | | | III | Mn-O | 0.34 |
| III-IV | $O_2$-La | 1.63 | 1.64 | 1.62 | 1.62 | III-IV | O-La | 1.53 |
| IV | La-O* | 0.22 | 0.23 | 0.23 | 0.24 | IV | La-O | 0.13 |
| IV-V | O-Mn | 1.94 | 1.90 | 1.95 | 1.89 | IV-V | O-O | 1.70 |
| V | Mn-$O_2$* | 0.11 | 0.08 | 0.14 | 0.10 | V | O-Mn | 0.23 |
| | | | | | | V | Mn-O | 0.36 |
| V-VI | $O_2$-La | 1.67 | 1.67 | 1.64 | 1.69 | V-VI | O-La | 1.54 |
| VI | La-O* | 0.26 | 0.21 | 0.28 | 0.17 | VI | La-O | 0.10 |
| VI-VII | O-Mn | 1.99 | 1.99 | 1.93 | 2.03 | VI-VII | O-O | 1.79 |
| VII | Mn-$O_2$* | 0.06 | 0.04 | 0.09 | 0.06 | VII | O-Mn | 0.23 |
| | | | | | | VII | Mn-O | 0.33 |
| VII-VIII | $O_2$-La | 1.59 | 1.59 | 1.57 | 1.62 | VII-VIII | O-La | 1.44 |
| VIII | La-O* | 0.44 | 0.43 | 0.43 | 0.45 | VIII | La-O | 0.38 |

* splitting of the bulk planes

Table 5. The relative Mn-O distances (Å) along the [001] direction for 8- (a) and 7-plane (b) slabs, and their differences for these two slabs.

| $MnO_2$ termination ||||||||||||||| 
|---|---|---|---|---|---|---|---|---|---|---|---|---|---|---|
| 8-plane ||||| 7-plane ||||| difference for 8 and -7 planes |||||
| FM | AAF | CAF | GAF | FM* | FM | AAF | CAF | GAF | FM* | FM | AAF | CAF | GAF | FM* |
| 2.12 | 2.12 | 2.12 | 2.13 | 2.09 | 2.10 | 2.12 | 2.11 | 2.12 | 2.07 | 0.02 | 0.00 | 0.01 | 0.01 | 0.02 |
| 1.88 | 1.88 | 1.87 | 1.87 | 1.85 | 1.88 | 1.85 | 1.87 | 1.85 | 1.84 | -0.01 | 0.03 | 0.00 | 0.02 | 0.00 |
| 2.00 | 2.01 | 2.00 | 2.02 | 2.00 | 1.95 | 1.94 | 1.94 | 1.93 | 1.95 | 0.05 | 0.08 | 0.06 | 0.10 | 0.05 |
| 1.94 | 1.90 | 1.95 | 1.89 | 1.93 | mirror plane ||||||||||
| 2.03 | 1.96 | 2.07 | 1.95 | 1.99 | 2.00 | 1.95 | 1.98 | 1.94 | 1.97 | 0.04 | 0.01 | 0.08 | 0.00 | 0.02 |
| 1.99 | 1.99 | 1.93 | 2.03 | 2.01 | 2.02 | 2.00 | 2.03 | 2.03 | 2.05 | -0.04 | -0.01 | -0.11 | 0.00 | -0.04 |
| 2.10 | 2.06 | 2.09 | 2.13 | 2.15 | 2.09 | 2.05 | 2.15 | 2.14 | 2.18 | 0.01 | 0.01 | -0.06 | -0.01 | -0.03 |
| LaO termination |||||||||||||||

* orthorhombic slab

Table 6. The effective atomic charges (a, b) and plane charges (c, d) for different magnetic states of 7-plane MnO$_2$-terminated (001) slab (a, c) and their deviations (b, d) from the bulk.

a)

| plane | atom | FM | AAF | CAF | GAF | FM** | atom | FM | AAF | CAF | GAF | FM** |
|---|---|---|---|---|---|---|---|---|---|---|---|---|
| I | Mn | 1.67 | 1.61 | 1.64 | 1.63 | 1.68 | O | -1.17 | -1.19 | -1.18 | -1.19 | -1.19 |
| II | La | 2.09 | 2.09 | 2.09 | 2.09 | 2.08 | O | -1.24 | -1.14 | -1.19 | -1.15 | -1.19 |
| III | Mn | 1.87 | 1.87 | 1.87 | 1.85 | 1.79 | O | -1.22 | -1.22 | -1.22 | -1.22 | -1.21 |
| IV | La | 2.15 | 2.14 | 2.14 | 2.14 | 2.07 | O | -1.33 | -1.34 | -1.35 | -1.34 | -1.21 |

b)

| plane | atom | FM | AAF | CAF | GAF | FM** | atom | FM | AAF | CAF | GAF | FM** |
|---|---|---|---|---|---|---|---|---|---|---|---|---|
| I | Mn | -0.18 | -0.23 | -0.21 | -0.22 | 0.00 | O | 0.15 | 0.13 | 0.14 | 0.14 | 0.05 |
| II | La | -0.04 | -0.04 | -0.04 | -0.04 | 0.01 | O | 0.09 | 0.18 | 0.13 | 0.17 | 0.06 |
| III | Mn | 0.02 | 0.02 | 0.02 | 0.01 | 0.11 | O | 0.10 | 0.10 | 0.11 | 0.11 | 0.05 |
| IV | La | 0.02 | 0.01 | 0.01 | 0.01 | 0.00 | O | -0.01 | -0.01 | -0.02 | -0.02 | 0.04 |

c)

| plane | | FM | AAF | CAF | GAF | FM** |
|---|---|---|---|---|---|---|
| I | MnO$_2$ | -0.68 | -0.78 | -0.73 | -0.75 | -0.71 |
| II | LaO | 0.85 | 0.95 | 0.90 | 0.94 | 0.89 |
| III | MnO$_2$ | -0.58 | -0.57 | -0.57 | -0.58 | -0.61 |
| IV | LaO | 0.82 | 0.80 | 0.80 | 0.79 | 0.86 |

d)

| plane | | FM | AAF | CAF | GAF | FM** |
|---|---|---|---|---|---|---|
| I | MnO$_2$ | 0.12 | 0.03 | 0.08 | 0.05 | 0.11 |
| II | LaO | 0.05 | 0.14 | 0.09 | 0.13 | 0.07 |
| III | MnO$_2$ | 0.22 | 0.23 | 0.24 | 0.22 | 0.21 |
| IV | LaO | 0.02 | 0.00 | -0.01 | -0.01 | 0.04 |

** orthorhombic slab

Table 7. The same as Table 6 for LaO-termination.

a)

| plane | atom | FM | AAF | CAF | GAF | FM** | atom | FM | AAF | CAF | GAF | FM** |
|---|---|---|---|---|---|---|---|---|---|---|---|---|
| I | La | 1.98 | 1.96 | 1.99 | 1.99 | 1.96 | O | -1.33 | -1.36 | -1.36 | -1.32 | -1.33 |
| II | Mn | 1.63 | 1.71 | 1.64 | 1.55 | 1.56 | O | -1.31 | -1.31 | -1.30 | -1.25 | -1.27 |
| III | La | 2.09 | 2.09 | 2.09 | 2.08 | 2.06 | O | -1.32 | -1.37 | -1.36 | -1.37 | -1.28 |
| IV | Mn | 1.76 | 1.79 | 1.79 | 1.70 | 1.64 | O | -1.31 | -1.31 | -1.30 | -1.27 | -1.26 |

b)

| plane | atom | FM | AAF | CAF | GAF | FM** | atom | FM | AAF | CAF | GAF | FM** |
|---|---|---|---|---|---|---|---|---|---|---|---|---|
| I | La | -0.15 | -0.17 | -0.14 | -0.14 | -0.11 | O | -0.01 | -0.03 | -0.04 | 0.00 | -0.08 |
| II | Mn | -0.21 | -0.13 | -0.21 | -0.30 | -0.11 | O | 0.02 | 0.01 | 0.02 | 0.07 | -0.03 |
| III | La | -0.04 | -0.04 | -0.04 | -0.05 | -0.01 | O | 0.01 | -0.04 | -0.03 | -0.05 | -0.03 |
| IV | Mn | -0.09 | -0.06 | -0.05 | -0.15 | -0.04 | O | 0.02 | 0.02 | 0.03 | 0.05 | -0.01 |

c)

| plane | | FM | AAF | CAF | GAF | FM** |
|---|---|---|---|---|---|---|
| I | LaO | 0.64 | 0.60 | 0.63 | 0.67 | 0.64 |
| II | MnO2 | -0.99 | -0.91 | -0.97 | -0.95 | -0.99 |
| III | LaO | 0.77 | 0.72 | 0.73 | 0.71 | 0.79 |
| IV | MnO2 | -0.85 | -0.82 | -0.80 | -0.85 | -0.88 |

d)

| plane | | FM | AAF | CAF | GAF | FM** |
|---|---|---|---|---|---|---|
| I | LaO | -0.16 | -0.21 | -0.17 | -0.13 | -0.18 |
| II | MnO2 | -0.18 | -0.10 | -0.16 | -0.15 | -0.16 |
| III | LaO | -0.03 | -0.08 | -0.07 | -0.10 | -0.03 |
| IV | MnO2 | -0.05 | -0.02 | 0.01 | -0.04 | -0.05 |

** orthorhombic slab

Table 8. The plane sequence for the (110) surface modelled using 8-plane (a), 7-plane LaMnO- (b), $O_2$- (c), and O- (d) terminated slabs.

| a) | b) | c) | d) |
|---|---|---|---|
| LaMnO | LaMnO | | |
| $O_2$ | $O_2$ | $O_2$ | O |
| LaMnO | LaMnO | LaMnO | LaMnO |
| $O_2$ | $O_2$ | $O_2$ | $O_2$ |
| LaMnO | LaMnO | LaMnO | LaMnO |
| $O_2$ | $O_2$ | $O_2$ | $O_2$ |
| LaMnO | LaMnO | LaMnO | LaMnO |
| $O_2$ | | $O_2$ | O |

Table 9. Cleavage energies of the FM (110) surface (in $eV/a_0^2$).

| slab | surface type | cubic | | orthorhombic | |
|---|---|---|---|---|---|
| | | unrelaxed | relaxed | unrelaxed | relaxed |
| nonstoichiometric | defectless | 2.60 | 1.69 | 2.70 | 1.75 |
| stoichiometric | defectless | 2.59 | 1.29 | 2.69 | 1.80 |
| | O-vac. symmetric | 1.59 | 0.75 | 1.68 | 0.96 |
| | O-vac. asymmetric | 1.60 | 0.49 | 1.68 | 0.76 |

Table 10. Atomic displacements (in % of $a_0 \sqrt{2}$) along the $y$ and $z$ axes for the defectless 7- and 8-plane (110) slabs. Positive (negative) sign means displacement outwards (inwards) the slab center.

| slab | 7-plane | | | | 8-plane | |
|---|---|---|---|---|---|---|
| termination | LaMnO | | $O_2$ | | | |
| atom | $\Delta y$ | $\Delta z$ | $\Delta y$ | $\Delta z$ | $\Delta y$ | $\Delta z$ |
| La | 0.00 | -6.43 | | | 0.00 | -6.10 |
| Mn | 0.00 | 3.36 | | | 0.00 | 3.40 |
| O | 0.00 | 1.98 | terminating plane | | 0.00 | 6.61 |
| $O_2$ | -0.34 | 0.25 | 1.02 | -4.98 | -1.45 | 3.26 |
| La | 0.00 | -0.47 | 0.00 | 2.56 | 0.00 | -4.77 |
| Mn | 0.00 | -0.43 | -0.01 | 2.15 | 0.00 | -0.17 |
| O | 0.00 | -0.31 | 0.00 | -3.49 | 0.00 | 1.13 |
| $O_2$ | 0.00 | 0.00 | 0.22 | -1.50 | -1.23 | 1.67 |
| La | Mirror plane | | 0.00 | 0.00 | 0.00 | -5.20 |
| Mn | | | 0.01 | 0.00 | 0.00 | -1.62 |
| O | | | 0.00 | 0.00 | 0.00 | 1.56 |
| $O_2$ | | | mirror plane | | -1.53 | 1.64 |
| La | | | | | 0.00 | -11.17 |
| Mn | | | | | 0.00 | -3.31 |
| O | | | | | 0.00 | 1.55 |
| $O_2$ | | | | | -1.32 | 2.48 |

Table 11. Atomic displacements along the *y* and *z* axes for the 7-plane O-terminated (110) slabs with oxygen vacancies in symmetrical and asymmetrical positions (Fig.4 )

| plane | atom | Symmetric | | asymmetric | |
|---|---|---|---|---|---|
| | | $\Delta y$ | $\Delta z$ | $\Delta y$ | $\Delta z$ |
| I | O | 9.98 | -10.26 | 9.09 | -9.32 |
| | O | -9.98 | -10.26 | -9.08 | -9.33 |
| II | La | 0.00 | 0.61 | 0.00 | -0.72 |
| | La | 0.00 | 0.61 | 0.00 | -0.72 |
| | Mn | 3.21 | 1.36 | 3.50 | 2.62 |
| | Mn | -3.22 | 1.36 | -3.50 | 2.62 |
| | O | 0.00 | -0.59 | 0.00 | 1.18 |
| | O | 0.00 | -0.59 | 0.00 | 1.18 |
| III | O | -5.51 | 10.57 | -0.96 | 3.87 |
| | O | 6.20 | -9.77 | 1.55 | -4.00 |
| | O | -6.21 | -9.76 | -1.55 | -4.00 |
| | O | 5.51 | 10.57 | 0.96 | 3.87 |
| IV | La | 0.00 | 0.00 | 0.00 | 0.00 |
| | La | 0.00 | 0.00 | 0.00 | 0.00 |
| | Mn | 0.01 | -0.01 | -0.19 | 0.00 |
| | Mn | 0.01 | 0.01 | 0.19 | 0.00 |
| | O | 0.00 | 0.00 | 0.00 | 0.00 |
| | O | 0.00 | 0.00 | 0.00 | 0.00 |

Table 12. The (111) cleavage energy in a cubic phase ($e$V/$a_0^2$)

| | | unrelaxed | relaxed |
|---|---|---|---|
| 7-plane | defectless | 2.77 | 2.68 |
| | Mn-vacancies | | 2.07 |
| 8-plane | defectless | 2.80 | 2.74 |

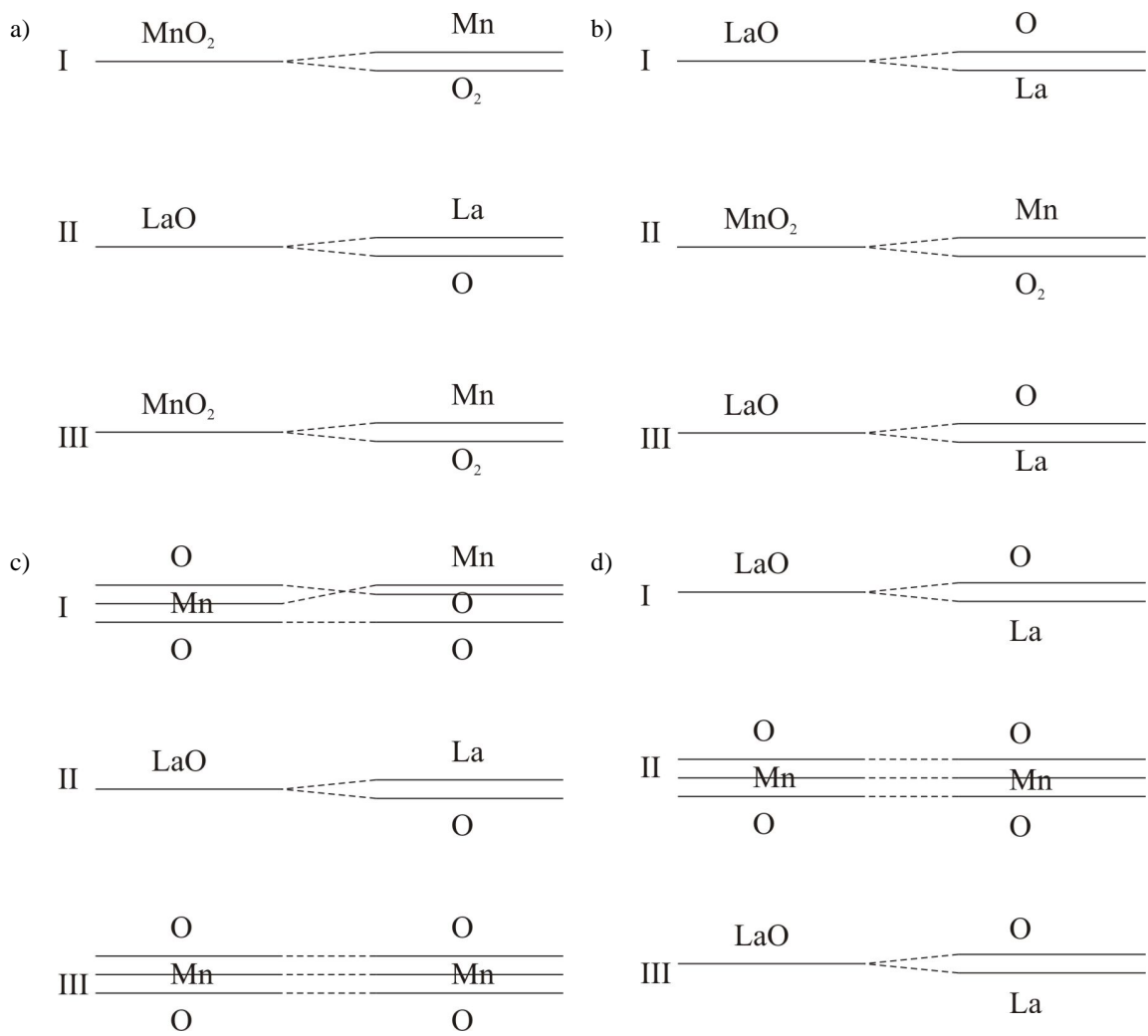

Figure 1. Splitting of the (001) surface planes (right side) with respect to the bulk (left side) in cubic (a), (b), and orthorhombic (c), (d) slabs for $MnO_2$- (a), (c) and LaO-termination (b), (d).

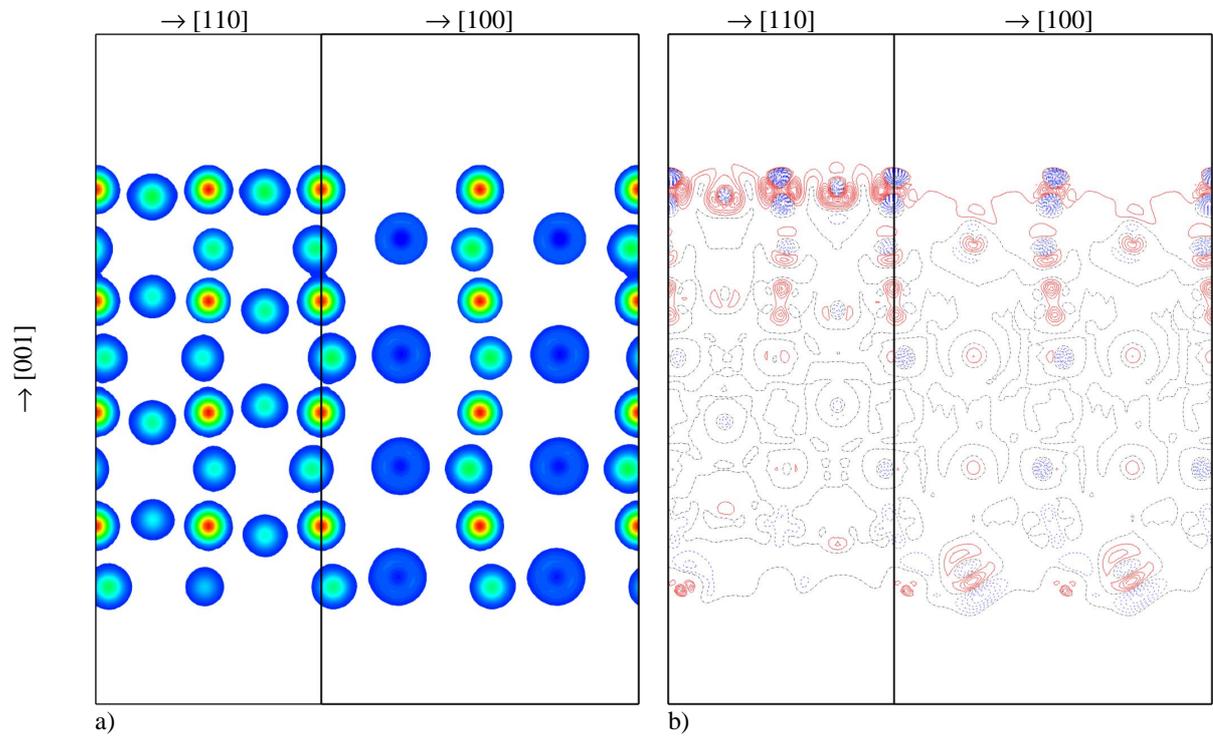

Figure 2. Total (a) and difference (b) electron density maps for 8-plane slab. Solid red and dash blue lines represent deficiency and excess of the electron charge, respectively. Density increment is 0.0125 e/Å$^3$. Dash-dot black line is the zero level.

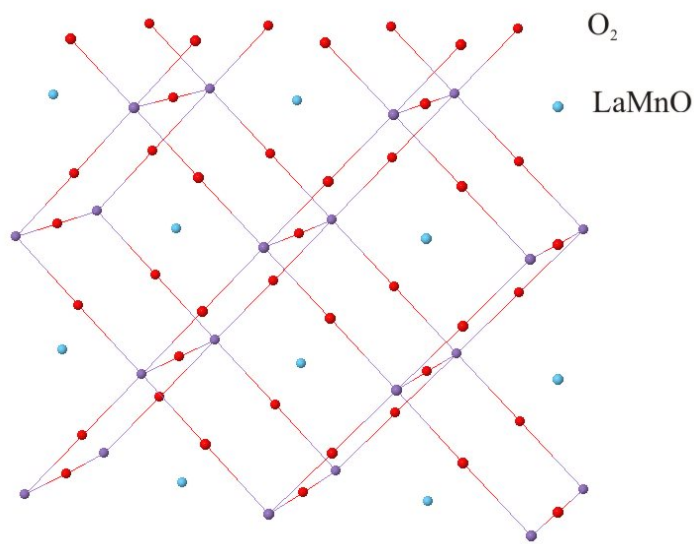

Figure 3. Schematic view of the (110) slab with even number of planes (no mirror plane). O$_2$ planes alternate with LaMnO ones along the direction normal to the surface.

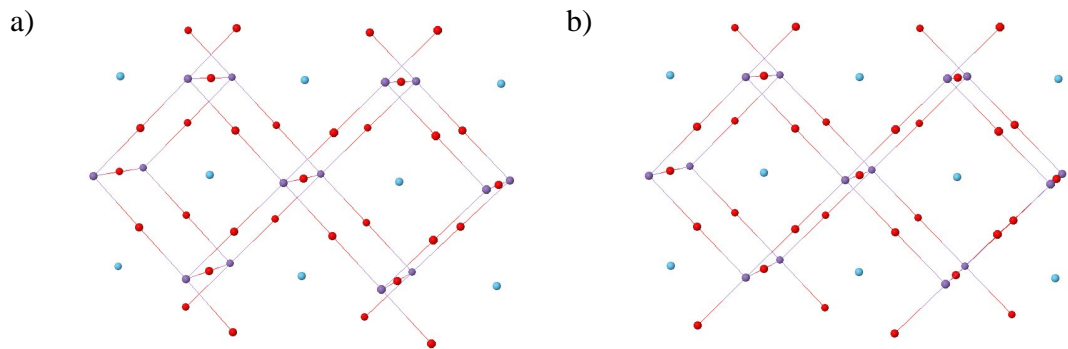

Figure 4. The 7-plane (110) O-terminated slab with half-filled surface oxygen planes. Terminating surfaces are symmetric (a) and antisymmetric (b) with respect to the mirror LaMnO plane.